\documentclass[aip,apl,reprint]{revtex4-2}

\usepackage{siunitx}
\usepackage{graphicx}
\usepackage{float}
\usepackage{mhchem}
\usepackage{dblfloatfix}
\usepackage{hyperref}

\begin{document}

\title{Role of polarity in the growth of cubic GaN within silicon inverted pyramids}

\author{David Lister}
\email[]{dal6@sfu.ca}
\affiliation{Simon Fraser University}

\author{Melissa Radford}
\affiliation{Simon Fraser University}

\author{Sara Fortin}
\affiliation{Simon Fraser University}

\author{Bilal Janjua}
\affiliation{Hyperlume}

\author{Mohsen Asad}
\affiliation{Hyperlume}

\author{Karen Kavanagh}
\affiliation{Simon Fraser University}

\author{Simon Watkins}
\email[]{simonw@sfu.ca}
\affiliation{Simon Fraser University}

\date{\today}

\begin{abstract}
A lack of spontaneous internal polarization makes cubic GaN (c-GaN) a well-suited material for emerging micro-LED-based short-range communication, where c-GaN promises increased speed over conventional hexagonal GaN (h-GaN). Although c-GaN is metastable, there are well-established methods for growing it in Si V- or U-grooves; the logical step is to truncate these grooves to wedges or inverted pyramids for small devices. There are limited reports of GaN grown in inverted pyramid templates, and the results are contradictory. To study this process, we perform selective area growth of GaN using organometallic vapor phase epitaxy (OMVPE) on Si inverted pyramidal templates and analyze our samples by cross-sectional TEM. We find that polarity is critical to understanding the growth of c-GaN in this four-fold geometry, in contrast to the growth in long grooves. This effect fits within the broader set of challenges of polar-on-nonpolar heteroepitaxy; the c-GaN inside the four-fold symmetric template has its symmetry reduced by polarity to be two-fold. In typical growth conditions---where the underlying h-GaN polarity is uniform---we find this implies that two h-GaN to c-GaN grain boundaries will have a polarity inversion. We observe two different structures at these inverting boundaries, including a previously unreported inversion domain boundary along the basal plane of the undoped h-GaN. These findings show that for small devices---such as micro-LEDs---the polarity-inverting interfaces must be prevented, for example by suppressing the growth of h-GaN on two facets of the template or by locally controlling the h-GaN polarity.

\end{abstract}

\maketitle 
Cubic gallium nitride (c-GaN) is an appealing material for optoelectronics because unlike hexagonal GaN (h-GaN), c-GaN's symmetry forbids spontaneous internal polarization. The reduced polarization significantly limits the quantum-confined Stark effect, suggesting increased optoelectronic device speed\cite{binks_cubic_2022} and emission efficiency\cite{dyer_efficiency_2024} resulting from greater electron and hole wavefunction overlap.

Metastability is the primary challenge of growing c-GaN. It is approximately 10 meV/atom less thermodynamically stable than h-GaN\cite{yeh_zinc-blendewurtzite_1992}, making growth by conventional means difficult. Among the multiple approaches that have been demonstrated,\cite{binks_cubic_2022} a well-known method involves nucleation of c-GaN by the merging of h-GaN crystals in V-grooves\cite{lee_spatial_2004,lee_nanoscale_2005} or U-grooves\cite{bayram_cubic_2014,liu_maximizing_2016} etched into Si(100).

The growth of small regions of high-quality c-GaN is of particular interest for micro-LEDs applied to short-range optical communications, where h-GaN has started to deliver performance improvements.\cite{benyahya_mosaic_2025} For small devices, a truncated V-groove in the form of an inverted pyramid or wedge makes a natural candidate. These have been recently explored as growth templates\cite{ansah-antwi_growth_2015,lee_initial_2019,khan_pyramidal_2022}; however, results are mixed. The literature is split between a result with no c-GaN formation,\cite{ansah-antwi_growth_2015} a claim that growth in a pyramid will improve c-GaN quality\cite{lee_initial_2019}, and a claim that pyramids improve the yield of c-GaN.\cite{khan_pyramidal_2022}

The growth process in a four-sided inverted pyramid or wedge is more complex than a V-groove, since merging of h-GaN can occur between adjacent as well as opposite h-GaN crystals within the template. In this letter, we identify facet polarity as an essential consideration for the growth of c-GaN in these templates. Surface polarity reduces the rotational symmetry of the c-GaN inside the pyramid to two-fold, conflicting with the four-fold symmetry of the template. When the h-GaN is grown with uniform polarity, this results in polarity-inverting interfaces between h-GaN and c-GaN and leads to significant defects.

For this study, a thermal \ce{SiO2} mask was grown on a 2$''$ Si(100) wafer. The wafer was patterned using a maskless aligner (MLA150) then etched using reactive ion etching. With \qty{3}{\micro\meter} square openings on the mask, the substrate was etched in \ce{KOH} to expose the Si\{111\} facets, forming an array of pyramidal cavities. The etch process undercut the mask by roughly \qty{0.5}{\micro\meter}, creating an overhang.

The sample was grown by OMVPE using a growth recipe developed for planar GaN on sapphire, but with an AlN buffer layer added to prevent meltback etching of the Si substrate. After an initial bake at \qty{1050}{\degreeCelsius}, \qty{60}{\nano\meter} of AlN was grown at \qty{750}{\degreeCelsius} followed by a thin low-temperature GaN buffer layer at \qty{470}{\degreeCelsius}. The main GaN layer was grown at \qty{1000}{\degreeCelsius} with a V/III ratio of 7000 (TMGa, \ce{NH3}, \ce{H2} carrier) to fill the template features.

Overall sample imaging was performed using scanning electron microscopy (SEM) and cross-sections were prepared using Ga focused ion beam (FIB) on a dual-beam instrument (FEI Helios) for transmission electron microscopy (TEM, Tecnai Osiris operated at 200 keV). TEM methods included bright field (BF), dark field (DF), selected area electron diffraction (SAED) and low-angle annular dark field (LAADF) scanning transmission electron microscopy (STEM).

\begin{figure}
    \centering
    \includegraphics[width=8.5cm]{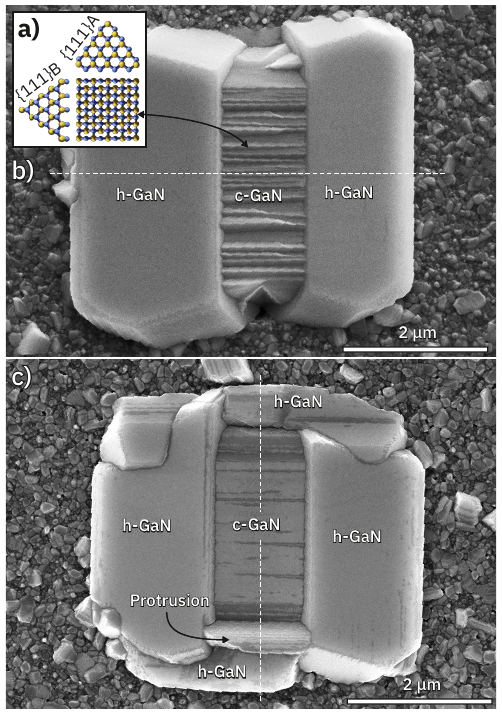}
    \caption{a) atomic model of the c-GaN in the same orientation, Ga yellow, N blue. b), c) SEM images of the growth on the templates.} 
    \label{fig:c-GaN_detail}
\end{figure}

The polarity of the c-GaN\{111\} facets is central to the analysis of the samples, so we will consider it in isolation first. Fig.~\ref{fig:c-GaN_detail} a) shows an atomic model pyramid of c-GaN oriented as in the SEM images. The bottom right is a plan view looking towards the flat top of the inverted pyramid. The side views are third-angle projections, where the polarity of the \{111\} facets is visible on the edges. Critically, the surface polarity of the c-GaN facets alternates between Ga- and N-polarity.

Examples of the growth in two pyramidal templates are shown in Fig.~\ref{fig:c-GaN_detail} b) and c). In both SEM images, the h-GaN in the pyramid has grown non-uniformly.\footnote{We attribute this non-uniformity to the mask overhang because other growths with the mask removed showed uniform h-GaN growth on all four facets.} The sample in Fig.~\ref{fig:c-GaN_detail} b) shows a case where h-GaN grew significantly on only the left and right facets of the pyramid, and the result is similar to that seen in a V-groove.\cite{lee_spatial_2004, lee_nanoscale_2005, lee_atomic-scale_2016, lee_initial_2019} The two h-GaN crystals merged to form a grain boundary, and then nucleated c-GaN. The c-GaN forms a wedge shape and fills the groove formed between the (0002) surfaces of the two h-GaN crystals. Because the h-GaN grew on only two opposite facets, this case is effectively a truncated V-groove.

\begin{figure}[h!]
    \centering
    \includegraphics[width=8.5cm]{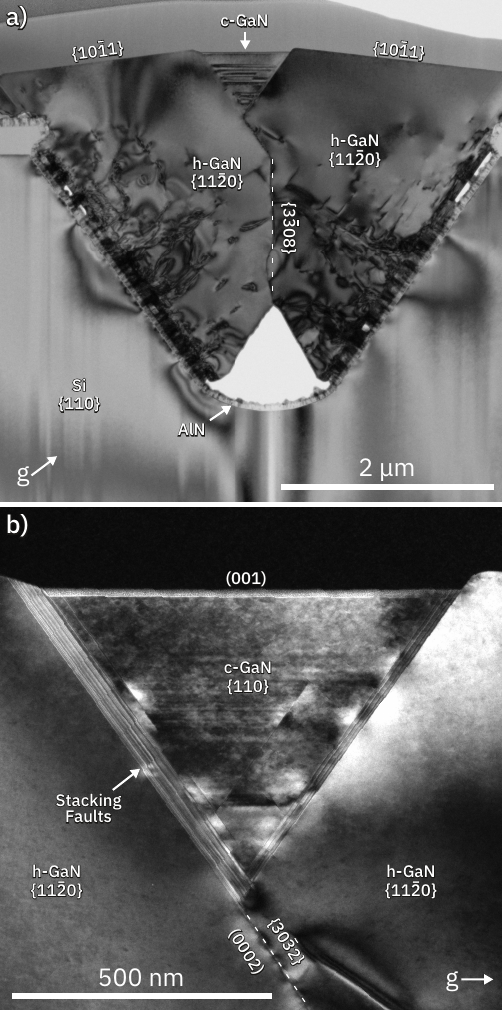}
    \caption{A TEM cross-section looking normal to the groove axis cut from Fig.~\ref{fig:c-GaN_detail} b). a) BF, $\mathbf{g}=(111)$. The h-GaN can be seen filling the Si cavity with a thin layer of AlN separating the GaN from the Si. A triangular region of c-GaN is visible between the two h-GaN crystals. b) Higher-magnification DF image, $\mathbf{g}=(220)$ of the cubic region.}
    \label{fig:S301_TEM_Normal}
\end{figure}

The dashed line across the middle of Fig.~\ref{fig:c-GaN_detail} b) shows the location of the TEM cross section shown in Fig.~\ref{fig:S301_TEM_Normal}. The two h-GaN crystals that formed the c-GaN are visible on the left and right sides of the TEM BF image in Fig.~\ref{fig:S301_TEM_Normal} a) in a two-beam condition, $\mathbf{g}=(111)$. The grain boundary in the middle is nominally a symmetric high-angle tilt boundary along \{3$\bar{3}$08\}; near the top it deflects left shortly before the c-GaN nucleated. Within the c-GaN itself, there are multiple horizontal lines. These are inclined stacking faults, with their contrast enhanced by the two-beam condition used for this image.

The transition from h-GaN to c-GaN is made clearer by DF, $\mathbf{g}=(220)$ shown in Fig.~\ref{fig:S301_TEM_Normal} b). Stacking faults normal to the cross-section have contrast enhanced by this beam condition and appear as the lines parallel to the hexagonal-cubic interfaces. The grain boundary at the bottom of the figure that formed immediately prior to c-GaN nucleation aligns with (0002) in the left h-GaN crystal and (3$\bar{3}$02) in the right crystal. Nucleation of the c-GaN appears to be preceded by a high density of stacking faults in the left h-GaN crystal. The causal pathway is not clear, however, because nucleation may have actually taken place at a different location along the groove out-of-plane to this cross section and the stacking faults may equivalently have formed to accommodate the growing c-GaN. The transitions between c-GaN and h-GaN are nevertheless planar and well ordered, with wetting of the h-GaN by the c-GaN on both sides. This transition is typical\cite{lee_spatial_2004,lee_nanoscale_2005, bayram_cubic_2014,liu_maximizing_2016} for h-GaN(0002) to c-GaN(111) interface when the polarity is continuous.

\begin{figure}[h!]
    \centering
    \includegraphics[width=8.5cm]{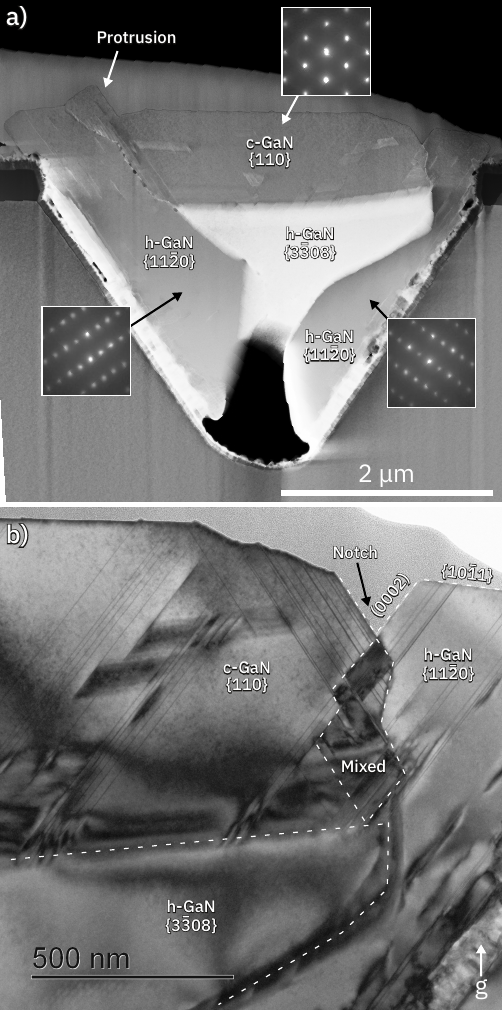}
    \caption{A cross-section parallel to the groove axis, along the dashed line from Fig.~\ref{fig:c-GaN_detail} c). a) LAADF STEM image showing the overall cross-section with SAED insets showing the h-GaN and central c-GaN regions. The bright region under the c-GaN comprises h-GaN from the out-of-plane crystals with the beam normal to their (3$\bar{3}$08) planes, giving significant diffraction contrast. b) TEM BF, $\mathbf{g}=(002)$ of the right interface between c-GaN and h-GaN showing many stacking faults and a complex structure. The dashed lines indicate the grain boundaries.}
    \label{fig:S301_TEM_Parallel}
\end{figure}

We now examine a more complex condition: the growth in Fig.~\ref{fig:c-GaN_detail} c) has h-GaN growing at different effective rates from all four facets. The left and right facets grew more quickly than the top and bottom, forming a groove similar to that in Fig.~\ref{fig:c-GaN_detail} b). The difference is that the ends of the groove are terminated by h-GaN grown from the top and bottom facets of the pyramid. As the model in Fig.~\ref{fig:c-GaN_detail} a) shows, the c-GaN wedge inside this structure will have alternating polarity of its \{111\} facets. The h-GaN in the inverted pyramid, however, grows with a nominally uniform polarity; with OMVPE this is typically Ga-polar.\cite{hellman_polarity_1998} This means that two of the four interfaces should have a polarity inversion and these will occur at the ends of the groove.

A cross-section along the dashed line is shown in Fig.~\ref{fig:S301_TEM_Parallel}, in which we examine the interfaces expected to have a polarity inversion based on the model in Fig.~\ref{fig:c-GaN_detail} a). The two h-GaN crystals at the end of the groove are seen merging with c-GaN in the LAADF STEM image of Fig.~\ref{fig:S301_TEM_Parallel} a). The bright region in the middle shows the two h-GaN crystals that merged to form the groove. The beam is normal to the (3$\bar{3}$08) plane in these crystals, providing high LAADF contrast. These crystals completely separate the left and right h-GaN crystals. Above the highly diffracting h-GaN, we see the c-GaN crystal and confirm its structure by SAED. The grain boundaries between the c-GaN and the left and right h-GaN crystals are the areas of interest.

The interface between c-GaN and h-GaN on the right of Fig.~\ref{fig:S301_TEM_Parallel} a), magnified in the BF image in Fig.~\ref{fig:S301_TEM_Parallel} b), shows a complex grain boundary that is distinctly different from the abrupt boundary seen in the prior cross-section. There is no abrupt junction; instead there is a region of mixed phase between the main crystals. There are also stacking faults that emanate from the grain boundary into the c-GaN rather than being parallel to the boundary. The c-GaN and h-GaN also appear to not wet readily and form a ``V'' notch where the grain boundary meets the sample surface. The lack of wetting indicates that the grain boundary energy exceeds the cost of the additional free surface of the notch. This high-energy interface stands in contrast to the polarity-continuous interface, which is structurally equivalent to a stacking fault and is correspondingly low in energy.

\begin{figure}[h!]
    \centering
    \includegraphics[width=8.5cm]{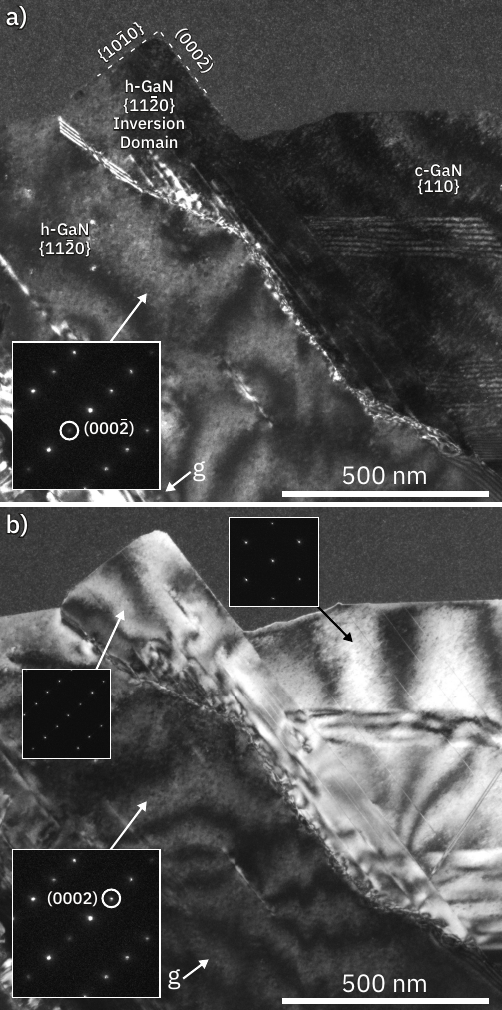}
    \caption{Dark-field image of the left interface from Fig.~\ref{fig:S301_TEM_Parallel}. a) $\mathbf{g}=(000\bar{2})$; b) $\mathbf{g}=(0002)$. The inset on the lower left of each is the SAED pattern of the left h-GaN crystal with the imaging spot highlighted. The two insets in b) show the SAED patterns of the inversion domain and the c-GaN respectively.}
    \label{fig:inversion_domain}
\end{figure}

The left interface has a very different character and instead contains an inversion domain, as seen in Fig.~\ref{fig:inversion_domain}; this domain corresponds to the protrusion labeled in Fig.~\ref{fig:c-GaN_detail} c). The DF images in Fig.~\ref{fig:inversion_domain} were taken with $\mathbf{g}=(000\bar{2})$ a) and $\mathbf{g}=(0002)$ b), giving polarity-sensitive contrast due to the breakdown of Friedel's law caused by the lack of inversion symmetry along the polar c-axis.\cite{serneels_friedels_1973} The inset SAED patterns show the bulk h-GaN and c-GaN as expected, and that the protrusion is h-GaN with the same orientation as the main h-GaN crystal on the left. In a) we see the c-GaN and the protrusion have matching contrast, while the main h-GaN crystal is brighter. In b) we see the opposite. This means the protrusion is an inversion domain with respect to the main h-GaN crystal. 

The inversion domain shows an interface with the c-GaN as abrupt as we see in the cross-section normal to the groove. The inversion domain also appears to form at the same height as the c-GaN within the groove and does not contact the AlN buffer layer. We believe this inversion domain is caused by the formation of c-GaN and not the reverse for the reasons to follow. Closer examination reveals a distinct and meandering grain boundary between the h-GaN and the inversion domain, roughly along the h-GaN basal plane. There is a well-known, relatively low-energy inversion domain boundary on the M-plane (11$\bar{1}$0) in undoped h-GaN,\cite{northrup_inversion_1996} but the boundary here lies approximately along the basal plane. This can occur with a high concentration of dopants such as Mg.\cite{northrup_magnesium_2003} To our knowledge, this is the first report of such an inversion domain without a dopant present. Due to the rarity of this defect, we believe it is most likely that the domain forms as a result of the polarity-inverting c-GaN/h-GaN interface.

Surface facet morphology provides an independent check on the polarity assignment. Relative facet kinetics that determine the crystal habit vary by growth technique and conditions. For OMVPE growth with a comparable V/III ratio, studies using kinetic Wulff plots show that the N-polar c-plane facet in h-GaN grows much slower than the other N-polar facets.\cite{sun_understanding_2008} Because slow-growing facets define the habit of convex crystals, the semipolar facets in the N-polar direction are small to non-existent. This means that the N-polar ($000\bar{2}$) abuts \{$10\bar{1}0$\} with a \qty{90}{\degree} corner and no facets between, while the Ga-polar (0002) has a large \{10$\bar{1}$1\} facet connecting it to \{$10\bar{1}0$\}. In our data, these are both visible. The Ga-polar surface is identified by a $\sim$\qty{120}{\degree} corner between (0002) and \{10$\bar{1}$1\} and is highlighted in Fig.~\ref{fig:S301_TEM_Parallel} b). The inversion domain shows the expected behavior for N-polar GaN, and Fig.~\ref{fig:inversion_domain} a) highlights the \qty{90}{\degree} corner. This is further evidence suggesting all four h-GaN crystals grew Ga-polar and the inversion domain is N-polar.

In summary, we find that if h-GaN polarity is uniform and all four crystals form, then the growth in an inverted pyramid will inevitably have two energetically unfavorable interfaces. Fundamentally, this is caused by the Si template having higher symmetry than the c-GaN that forms. This problem is similar to the antiphase disorder or frustration found when growing GaAs on Ge\cite{lazzarini_antiphase_2000} or Si\cite{kroemer_polar--nonpolar_1987} and can be seen within the broader set of challenges of polar-on-nonpolar heteroepitaxy.

Our data show that there are at least two different ways that the h-GaN to c-GaN interface with polarity inversion can be resolved, and one of them forms a basal-plane inversion domain. This suggests that for small devices, inverted pyramids or wedges pose a challenge, and that a means of controlling the polarity-inverting interfaces is likely needed. We also see that when there are no crystals terminating the groove, the c-GaN fills it up to the ends, suggesting at least one path forward is to reliably prevent h-GaN growth on these terminal facets.

To generalize, within a zincblende crystal there are two possible dihedral angles between non-parallel \{111\} facets. The A-A and B-B planes intersect at a dihedral angle of \qty{70.5}{\degree} while the angle between A and B facets is \qty{109.5}{\degree}. These correspond to angles between opposite (\qty{70.5}{\degree}) and adjacent (\qty{109.5}{\degree}) facets in the pyramid. When the h-GaN polarity is uniform, only merging at the \qty{70.5}{\degree} condition will produce a low-energy interface between c-GaN and both h-GaN crystals. A V- or U-groove on Si(100) presents a \qty{70.5}{\degree} dihedral angle, explaining the success of this geometry for growing c-GaN.

The generalization also leads to a prediction. We find c-GaN preferentially nucleates when opposing facets merge\footnote{More detailed analysis is in progress for a forthcoming publication.} during growth. This is because the h-GaN polarity is uniform, and the polarity inversion that results from c-GaN bridging two adjacent facets is energetically costly and probably inhibits c-GaN formation. If the h-GaN polarity on two adjacent facets differs, this situation is reversed. Merging between adjacent facets of opposite polarity at a dihedral angle of \qty{109.5}{\degree} creates an arrangement in which c-GaN can bridge these two h-GaN crystals without an inverting interface, and this grain boundary may nucleate c-GaN as it evolves.\footnote{This grain boundary is different from the one between two opposite crystals, so further experiments are needed to support this prediction.} Interlayers have been shown to be effective at polarity control on \ce{Al2O3}\cite{mita_fabrication_2009} and Si(111)\cite{brubaker_polarity-controlled_2016}. The pyramidal geometry also allows for shadowed deposition of this interlayer or a masking layer. Combined, these two concepts could be used to realize pyramidal or wedge geometries where all four facets have the correct polarity for h-GaN to merge with c-GaN.

\begin{acknowledgments}
The authors gratefully acknowledge funding from the Natural Sciences and Engineering Research Council of Canada (NSERC). DL is especially grateful for support from an NSERC Canada Graduate Research Scholarship—Doctoral (CGRS D), and to Hyperlume (now Credo Semiconductor) for their valuable collaboration and financial support through a MITACS Accelerate award. This work made use of the 4D LABS core facility at Simon Fraser University (SFU).
\end{acknowledgments}

\bibliography{references}

\end{document}